\documentclass[%
 aip,
 amsmath,amssymb,
 reprint,%
]{revtex4-1}

\usepackage{graphicx}
\usepackage{dcolumn}
\usepackage{bm}

\usepackage[utf8]{inputenc}
\usepackage[T1]{fontenc}
\usepackage{mathptmx}
\usepackage{etoolbox}
\usepackage[version=4]{mhchem}

\makeatletter
\def\@email#1#2{%
 \endgroup
 \patchcmd{\titleblock@produce}
  {\frontmatter@RRAPformat}
  {\frontmatter@RRAPformat{\produce@RRAP{*#1\href{mailto:#2}{#2}}}\frontmatter@RRAPformat}
  {}{}
}%
\makeatother
\begin{document}

\preprint{AIP/123-QED}

\title{Inverted Hanle spin precession induced magnetoresistance in chiral/semiconductor systems}
\author{S.H. Tirion}
 \affiliation{Zernike Institute for Advanced Materials, University of Groningen, NL-9747AG Groningen, The Netherlands}
  \email{s.h.tirion@rug.nl.}
\author{B.J. van Wees}%
\affiliation{Zernike Institute for Advanced Materials, University of Groningen, NL-9747AG Groningen, The Netherlands
}%

\date{\today}

\begin{abstract}
In the past decade, chiral materials have drawn significant attention because it is widely claimed that they can act as spin injectors/detectors due to the chirality-induced spin selectivity (CISS) effect. Nevertheless, the microscopic origin of this effect is not understood, and there is an intensive discussion about the manifestation of the magnetoresistance that is generated between a chiral system and a ferromagnet. Hanle spin precession measurements can unambiguously prove the injection and detection of a spin accumulation in a non-magnetic material, as was shown with traditional ferromagnetic injectors/detectors. Here, we analyze in detail the Hanle spin precession-induced magnetoresistance and find that it is inverted as compared to the ferromagnetic case. We explicitly model the spin injection and detection by both a chiral system and a ferromagnetic system, as well as the spin transport in a semiconductor, for a general set of (spin) transport parameters that cover the relevant experimental regime. For all sets of parameters, we find that the Hanle signals for a chiral system and ferromagnet are each other's opposites. 
\end{abstract}

\maketitle

Spintronic devices rely on the efficient interconversion of optical or electrical signals into spin signals and vice versa to inject and detect electron spins. Hanle spin precession measurements are a useful tool to confirm the presence of a spin accumulation, for instance, semiconductors \cite{Spiesser2019,Dash2009,Lou2006,Kamerbeek2015}, systems low dimensional systems such as graphene \cite{Tombros2007,Hu2020} and metallic systems. \cite{Velez2016,Jedema2002} Increased interest was generated in Hanle spin precession measurements with chiral materials both theoretically, \cite{Yang2020a,Yang2021} as well as recently experimentally. \cite{Sun2024,Liu2024} This is due to the chirality-induced spin selectivity (CISS) effect, \cite{Bloom2024,Firouzeh2024,Evers2022} which is studied in numerous materials, including organic molecules, \cite{Mishra2020,Liu2020,Mondal2021,Safari2023,Firouzeh2023,Singh2024} but also inorganic chiral systems. \cite{Lu2019,Inui2020,Bian2023} Furthermore, the rise of atomically flat two-dimensional materials and the ability to controllably stack them with a twist angle opens up new avenues to investigate the connection between chirality and (spin-dependent) electron transport. \cite{Menichetti2024}

Electron transport experiments in chiral materials contacted by a ferromagnetic electrode have shown a surprisingly large two-terminal magnetoresistance (MR), which is commonly interpreted in terms of spin(-dependent) transport in the chiral system, \cite{Michaeli2019a,Das2024,Moharana2024} in analogy to spin valves comprising of two ferromagnets such as in the giant magnetoresistance (GMR) or the tunnel magnetoresistance (TMR). However, the origin of the CISS MR is poorly understood. \cite{Evers2022} Furthermore, a number of recent reports also clearly show an MR in the linear response regime (around zero bias),\cite{Liu2020,Mondal2021,AlBustami2022a,Bian2023,Singh2024} which contradicts the Onsager reciprocity relations for two terminal measurements. \cite{Yang2019,Yang2020a}

\begin{figure}[h!]
\includegraphics[width=0.48\textwidth]{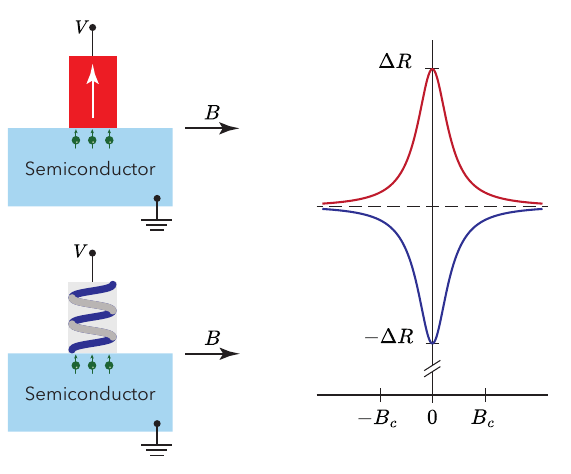}
\caption{\label{fig:main} Illustration of the crucial difference between Hanle spin precession for a chiral system and a ferromagnet as injector and detector coupled to a semiconductor. In the ferromagnetic case, the generation of a spin accumulation \textit{increases} the total resistance at $B = 0$. In contrast, for a chiral system as the spin injector and detector, the inverse effect occurs; the total resistance \textit{decreases} when a spin accumulation is built up at $B = 0$.}
\end{figure}

\begin{figure*}
\centering
\includegraphics[width=0.98\textwidth]{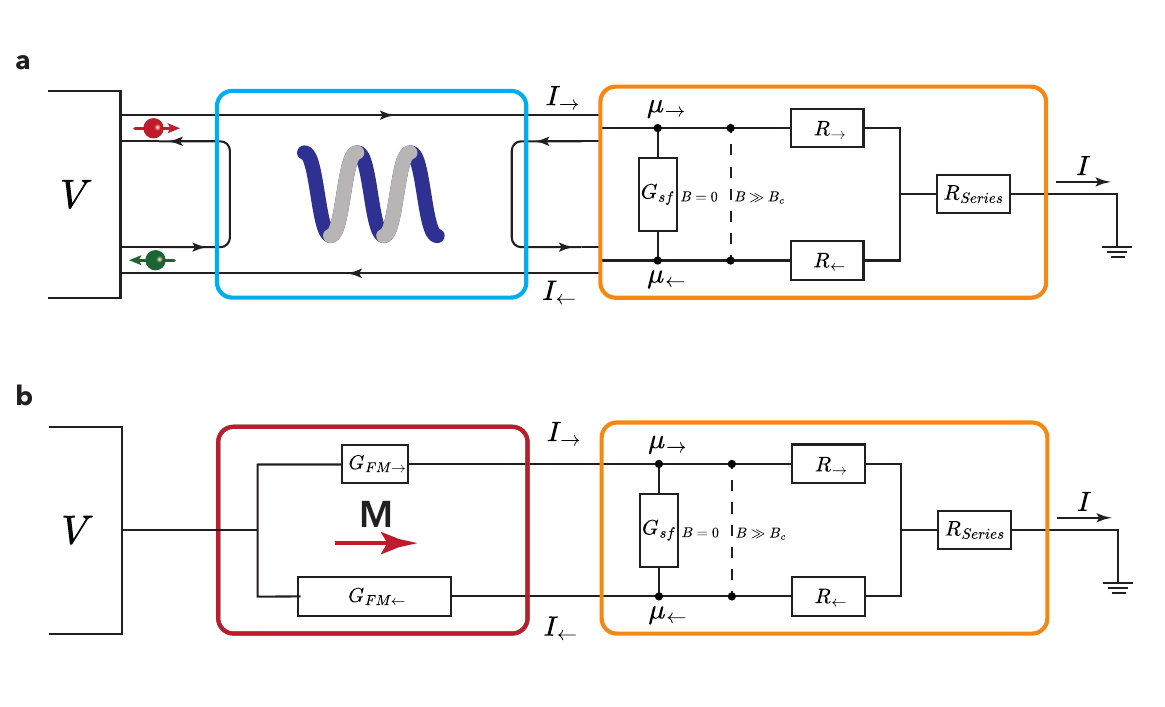}
\caption{\label{fig:Hanle_chiral} (a) Schematic illustration of the equivalent circuit for a chiral element coupled to a semiconductor for Hanle spin precession measurements. (b) for a system where a ferromagnetic material is connected to a semiconductor via spin-dependent tunneling. Three parts of this circuit can be distinguished. (i) The chiral element/ferromagnet that acts as injector and detector. (ii) The interface region of the chiral element/ferromagnet and the semiconductor, where we can calculate the spin accumulation, $\mu_s = \mu_{\rightarrow} - \mu_{\leftarrow}$. $G_{sf}$, the spin relaxation conductance, models the spin relaxation in the semiconductor. For $B \gg B_c$, Hanle spin precession short circuits the two spin channels (indicated with the dashed line). (iii) In the semiconductor, the separate spin species propagate over a distance $\lambda_{sf}$ via $R_{\rightarrow}$ and $R_{\leftarrow}$. Beyond $\lambda_{sf}$ the current passes a series resistor $R_{series}$. On the left-hand side, we apply a voltage $V$, and the righthand side of the circuit is held at ground potential. Note that the convention for the sign of the currents $I$, $I_{\rightarrow}$ and $I_{\leftarrow}$ is positive for currents flowing from left to right.}
\end{figure*}

Recently, alternative mechanisms to generate an MR in a chiral system coupled to a ferromagnet were proposed, \cite{Tirion2024,xiao2023nonreciprocal} based on the modification of the electron transport barrier and bandstructure by either magnetization or chirality reversal, which alters the charge transport and gives rise to an MR without requiring spin (dependent) transport. The experimental time scales on which the CISS MR is observed suggest a currently not understood persistent/equilibrium effect. Xiao et al. \cite{xiao2023nonreciprocal} have proposed that a bias current-induced charge trapping generates a long-lived state that modifies the transport barrier. However, the physical origin of these proposed chiral-magnetic electrostatic effects remains elusive for the moment.

It is therefore crucial to identify the mechanism that generates the linear response MR in chiral systems coupled to a ferromagnet. Hanle spin precession with a pure chiral system offers an excellent experimental approach because no ferromagnetic element is required, which eliminates possible contributions of an electrostatic transport barrier modification.

In this letter, we explicitly study the characteristics of Hanle spin precession experiments for chiral systems. We describe the spin and charge transport in a chiral element, which we connect to a semiconductor in which a spin accumulation is generated. For this system, we develop the equivalent circuit and obtain a universal result of Hanle spin precession measurements with any chiral system as injector and detector of spins. We find that the total resistance is always \textit{lowered} when spin accumulation is built up at the interface of the semiconductor with the chiral system. Upon the application of a magnetic field ($B \gg B_c$), the total resistance is \textit{increased} as the spins dephase. This result is inverted compared to the signal for the conventional configuration, where a ferromagnet is used as the spin injector and detector. This makes Hanle spin precession measurements with chiral systems a powerful tool to confirm spin injection and detection by chiral materials. We stress that our result is valid for a general chiral system.

Hanle spin precession with a chiral system on a semiconductor was recently studied experimentally by \citeauthor{Liu2024}, \cite{Liu2024} and it is claimed that Hanle spin precession was observed. However, our theoretical results predict an opposite sign compared to \citeauthor{Liu2024} In addition, the experimental lineshape that was observed is not of the typical Lorentzian form. This is also not compatible with our description of Hanle spin precession with a chiral system.

Hanle spin precession measurements have proven instrumental in determining if electronic spins are injected in a nonmagnetic channel and lead to a spin accumulation. \cite{Jedema2002,Dash2009,Spiesser2019, Hu2020} Three processes are required for Hanle spin precession measurements. The first is the electrical injection of spins by a spin-polarizing contact, creating a difference in the electrochemical potentials of the two spin states, $\mu_s = \mu_{\rightarrow} - \mu_{\leftarrow} $ (spin accumulation), which is proportional to the spin polarization, $P$, of the contact at the interface with the semiconductor. The second process is the reciprocal effect, which facilitates the electrical detection of the spin accumulation. The third process is the controlled reduction of $\mu_s$ by an external magnetic field $B$ perpendicular to the polarization direction of $\mu_s$, known as the Hanle spin precession effect. The external magnetic field causes the spin to precess at a Larmor frequency of $\omega_L=g\mu_BB/ \hbar $ where $g$ in the Land\'e $g$-factor, $\mu_B$ is the Bohr magneton and $\hbar$ is the reduced Planck’s constant. As a result, the spin accumulation decays with a Lorentzian line shape
\begin{equation}
	\mu_s(B) = \mu_s(B=0) \frac{1}{1+(\omega_L \tau_{sf})^2},
	\label{eq:Hanle_Lor}
\end{equation}
where $\tau_{sf}$ is the spin relaxation time. This is under the assumption of the absence of diffusion of the spins. The critical field is defined as $B_c = \hbar / g\mu_B \tau_{sf} $. We note that deviations from a Lorentzian line shape are possible, \cite{Kamerbeek2015,Das2018} but this requires the system to be biased in the nonlinear regime.

To describe Hanle spin precession measurements, we consider a semiconductor connected to a chiral element  (spin injector and detector) depicted in FIG.~\ref{fig:Hanle_chiral}(a) and a ferromagnet in FIG.~\ref{fig:Hanle_chiral}(b). The chiral element is biased with $\mu = -eV$. At the interface with the semiconductor, we define the spin accumulation $ \mu_s  = (\mu_{\rightarrow}-\mu_{\leftarrow})/2 $, as well as the charge electrochemical potential $ \mu_c  = (\mu_{\rightarrow}+\mu_{\leftarrow})/2 $. And describe the charge current $I = I_{\rightarrow} + I_{\leftarrow}$ and the spin current $I_{s} = I_{\rightarrow} - I_{\leftarrow}$, both in electrical units, at the interface with the semiconductor, with the spin-dependent transport matrix developed by Yang et al. \cite{Yang2019,Yang2020a} for a chiral system.

\begin{eqnarray}
{\begin{bmatrix}
I \\
I_{s}\end{bmatrix}} = -\frac{1}{e} {\begin{bmatrix}
G & GP_{ciss} \\
GP_{ciss} & -G^* \\
\end{bmatrix}}
{\begin{bmatrix}
\mu - \mu_c  \\
 \mu_s \\
\end{bmatrix}}.
\label{eq:transport_matrix_chiral}
\end{eqnarray}
We note that this matrix is the reduced form of the 3x3 matrix that was obtained in Ref.~\cite{Yang2020a} since there is no spin accumulation assumed on the left side of the chiral system and the spin current on the left side of the chiral element is not relevant for this analysis. This is the most general description of spin-dependent transport in a (nonmagnetic) chiral element, based on the symmetries of spin and charge transport. Note that this model does not make assumtions about the microscopic origin of the spin-dependent transport in the chiral element. Similar results were recently obtained by Korytár et al.\cite{Korytár2024} for a unitary model.

The matrix has three parameters. $G$ is the charge conductance of the chiral system. $P_{ciss}$ describes the CISS-induced spin polarization of the current. We note that the off-diagonal elements of the transport matrix are identical, as required by the Onsager reciprocity relations. \cite{Jacquod2012} The parameter $G^*$ describes the spin current that flows into the chiral element as a consequence of the built up spin accumulation. Spin-dependent transmission in a chiral system always requires spin-flip reflection processes. \cite{Yang2019,Wolf2022} This yields the condition that $G^* \geq G$. In the following analysis, we use $G^* = G$, so we ignore additional spin relaxation processes in the chiral element. Furthermore, we assume that the applied magnetic field does not modify the spin-dependent parameters in Eq.~(\ref{eq:transport_matrix_chiral}) of the chiral element.

\begin{figure}
\includegraphics[width=0.48\textwidth]{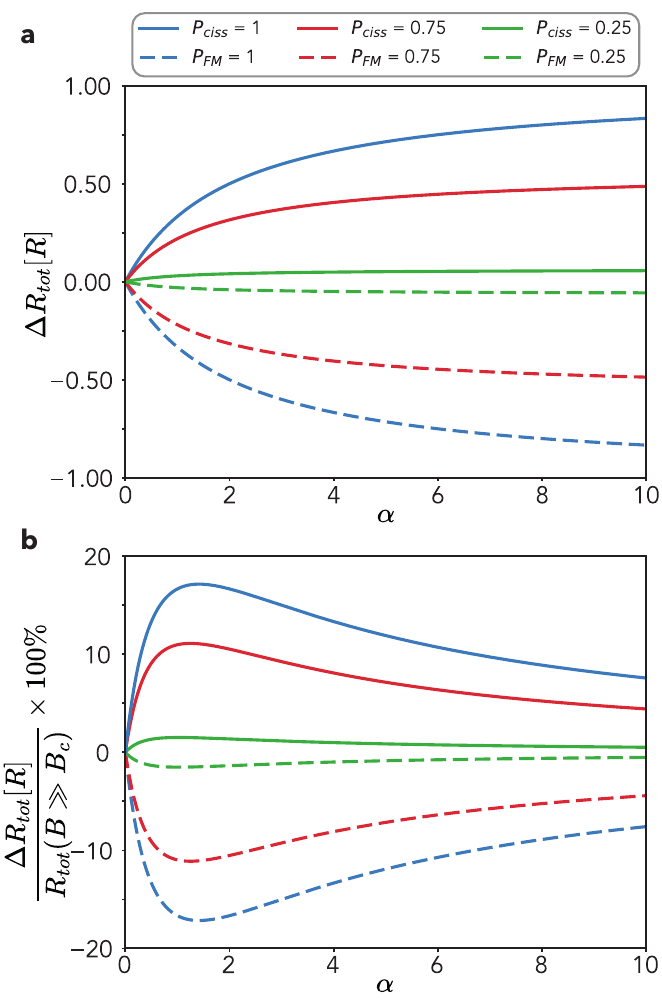}
\caption{\label{fig:plot_alpha_dep}(a) Hanle magnetoresistance $\Delta R_{tot} = R_{tot}(B \gg B_c) - R_{tot}(B=0)$ in units of $R$ as a function of $\alpha = \frac{1}{RG}$ for a chiral element (solid) and a ferromagnet (dashed). (b) $\Delta R_{tot}$ normalized with respect to $R_{tot}(B \gg B_c)$ which has a maximum at $\alpha_{max} = \sqrt{P^2+1}$.}
\end{figure}

The right side of the chiral element is coupled to a semiconductor, which we describe with the equivalent circuit shown in FIG.~\ref{fig:Hanle_chiral}. $G_{sf}$, between the $\mu_{\rightarrow}$ and $\mu_{\leftarrow}$ channels, describes the spin-relaxation processes in the semiconductor. These are dependent on the spin relaxation time, $\tau_{sf}$, as well as the volume $Vol$ with the typical dependence of $G_{sf} \sim \frac{Vol}{\tau_{sf}}$. We model the transport of spins with two spin channels, $R_{\rightarrow}=R_{\leftarrow}=2R$, which typically takes place over the spin relaxation length, $\lambda_{sf}$. Beyond the spin relaxation length, we describe the transport by a series resistance, $R_{Series}$. We note that this is the most general description of spin transport and relaxation in a semiconductor, which is, in practice, strongly dependent on the dimensionality of the semiconductor system and the character of the spin and charge flows, which we do not describe in detail in this work. For the remainder of this discussion, we take $ R_{Series} \ll R $, such that we can ignore $R_{Series}$.

We first consider the case without spin relaxation in the semiconductor when $G_{sf} =0$. Note that even though we have not explicitly assumed spin relaxation in the semiconductor, spin relaxation still occurs via the channels formed by $R_{\rightarrow}$ and $R_{\leftarrow}$, such that the charge and spin current in the semiconductor can be described by,

\begin{subequations}
\begin{alignat}{4}
& I = -\frac{1}{e} \frac{1}{2R} ( \mu_{\rightarrow} + \mu_{\leftarrow}) \\
& I_s = -\frac{1}{e} \frac{1}{2R} ( \mu_{\rightarrow} - \mu_{\leftarrow})
\end{alignat}
\label{eq:charge_spin_semi}
\end{subequations}

For the case when $B=0$, and using Eq.~(\ref{eq:transport_matrix_chiral}) and (\ref{eq:charge_spin_semi}), we obtain the total resistance of the circuit of the chiral element coupled to the semiconductor,

\begin{eqnarray}
R_{tot}(B=0) = R \frac{(\alpha+1)^2+P_{ciss}^2}{\alpha+1+P_{ciss}^2}.
\label{eq:R_tot_ciss} 
\end{eqnarray}
Here we introduce the parameter $\alpha = \frac{1}{RG}$. Upon the application of a magnetic field perpendicular to the direction of spin accumulation, the spins precess and dephase to zero, Eq.~(\ref{eq:Hanle_Lor}). We consider the case when the Hanle pressession has fully reduced the spin accumulation to zero, ($B \gg B_c $), modeled by a short circuit between the two spin channels. The total resistance is then given by the Ohmic addition of the chiral element and the semiconductor,

\begin{figure}
\includegraphics[width=0.48\textwidth]{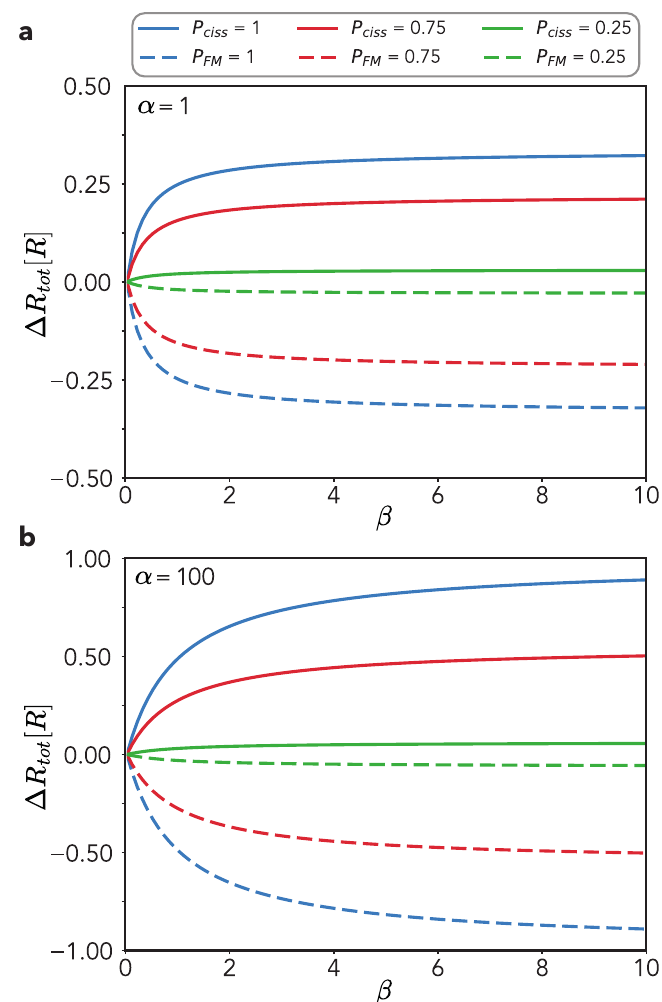}
\caption{\label{fig:plot_beta_dep} (a) Hanle magnetoresistance in units of $R$ as a function of $\beta = \frac{1}{RG_{sf}}$ for a chiral element (solid) and a ferromagnet (dashed)for $\alpha = 1$ and (b) for $\alpha=100$. }
\end{figure}

\begin{eqnarray}
R_{tot}(B \gg B_c) = R (1+\alpha).
\label{eq:R_tot_inf} 
\end{eqnarray}
Here we note that for any value of $\alpha$, Eq.~(\ref{eq:R_tot_ciss}) is smaller than Eq.~(\ref{eq:R_tot_inf}). This shows that the total resistance of a chiral element coupled to a semiconductor is always \textit{reduced} when a spin accumulation is built up. This reduction of the total resistance is destroyed due to dephasing of the spins by an applied field. We define the Hanle magnetoresistance $\Delta R_{tot}$ as the difference between the total resistance with magnetic field and without magnetic field, $\Delta R_{tot} = R_{tot}(B \gg B_c) - R_{tot}(B=0)$, plotted in FIG.~\ref{fig:plot_alpha_dep}(a) in units of $R$ for different values of $P_{ciss}$. We emphasize that this is a general result that is valid for all chiral systems and, therefore, forms a key signature of the Hanle spin precession measurements with chiral systems as the spin injector and detector. Generally, the resistance reduction scales with $P_{ciss}^2$, the product of the injection and detection efficiencies. For a chiral system with $P_{ciss} = 1$, the Hanle magnetoresistance, $\Delta R_{tot} \rightarrow R$ when $\alpha$ becomes large, such that the resistance of the semiconductor gets completely suppressed when the spins propagate via separate spin channels.

To quantity the Hanle magnetoresistance, we normalize $\Delta R_{tot}$ with respect to $R_{tot}(B \gg B_c)$ and plot it in FIG.~\ref{fig:plot_alpha_dep}(b) for different values of $P_{ciss}$. We find that the change in the normalized resistance reaches a maximum at $\alpha_{max} = \sqrt{P^2+1}$, which for a perfect chiral system ($P_{ciss} = 1$) gives a change in normalized resistance of $3-\sqrt{2} \approx 17\%$.

It is relevant to compare these results to a conventional Hanle spin precession circuit,\cite{Dash2009,Lou2006,Jansen2012} which uses a single ferromagnetic electrode as the injector and detector as illustrated in FIG.~\ref{fig:Hanle_chiral}(b). The spin injection by the ferromagnet we describe as spin-dependent tunneling with two different conductivities $G_{FM \rightarrow}$ and $G_{FM \leftarrow}$. \cite{Fert2001,Jansen2012} The total charge conductivity of the ferromagnet is $G_{FM} = G_{FM \rightarrow} + G_{FM \leftarrow}$. We introduce $P_{FM} = (G_{FM \rightarrow} - G_{FM \leftarrow})/(G_{FM \rightarrow} + G_{FM \leftarrow})$ to describe the spin polarization of the current. A similar analysis can be applied to obtain the total resistance of the circuit of a ferromagnet coupled to a semiconductor. We find that

\begin{eqnarray}
R_{tot}(B=0) = R \frac{(\alpha+1)^2 + P_{FM}^2(2\alpha +1)}{\alpha+1+P_{FM}^2},
\label{eq:R_tot_fm} 
\end{eqnarray}
where $\alpha = \frac{1}{R G_{FM}}$. Similarly to the case of the chiral system, we plot $\Delta R_{tot}$ as a function of $\alpha$ in FIG.~\ref{fig:plot_alpha_dep}(a) for different values of $P_{FM}$. Here, we obtain the familiar result where the total resistance is \textit{increased} when a spin accumulation is built up by a ferromagnetic contact. \cite{Lou2006,Dash2009,Jansen2012} The consequence of the Hanle spin precession in a ferromagnetic case is the reduction of the total resistance. When we compare the $\alpha$-dependence in FIG.~\ref{fig:plot_alpha_dep} for a chiral element and the ferromagnet, we find that the two are each other's opposites, which is an interesting result given the very different microscopic origins of spin-dependent transport in a ferromagnet and a chiral element. For both cases, we note that the sign of the signal does not dependent on the (sign of) chirality nor on the spin of the spin polarization of the ferromagnet

Thus far, we have taken $G_{sf}=0$ such that we could ignore spin-relaxation processes in the semiconductor. However, in realistic systems, this is not the case; therefore, we extend Eq.~(\ref{eq:R_tot_ciss}) and Eq.~(\ref{eq:R_tot_fm}) to include spin-relaxation processes in the semiconductor. We introduce $\beta = \frac{1}{RG_{sf}}$ and for the chiral system we find that

\begin{eqnarray}
R_{tot}(B=0) = R \frac{(\alpha+1)(\alpha \beta + \alpha + \beta)+\beta P_{ciss}^2}{\alpha(\beta+1)+\beta(P_{ciss}^2+1)}.
\label{eq:beta_dep_ciss} 
\end{eqnarray}
Using a similar analysis for the ferromagnetic system, we find that

\begin{eqnarray}
R_{tot}(B=0) = R \frac{(\alpha+1)(\alpha \beta + \alpha + \beta) + (2\alpha+1)\beta P_{FM}^2}{\alpha(\beta+1)+\beta(P_{FM}^2+1)}.
\label{eq:beta_dep_FM} 
\end{eqnarray}
In FIG.~\ref{fig:plot_beta_dep}(a) we plot $\Delta R_{tot}$ as function of $\beta$ for $\alpha = 1$ and in FIG.~\ref{fig:plot_beta_dep}(b) for $\alpha = 100$. We first consider the limiting cases for $\beta$ and find that when $\beta =0$ both $R_{tot}(B=0)$ in Eq.~(\ref{eq:beta_dep_ciss}) aswell as in Eq.~(\ref{eq:beta_dep_FM}) reduce to the result obtained for $R_{tot}(B \gg B_c)$. Physically, this can be understood as the increase of spin relaxation events ($G_{sf}$ becomes large) in the semiconductor, which reduces the spin accumulation in the semiconductor. In the limit when $\beta \rightarrow \infty$ Eq.~(\ref{eq:beta_dep_ciss}) reduces to Eq.~(\ref{eq:R_tot_ciss}) and Eq.~(\ref{eq:beta_dep_FM}) reduces to Eq.~(\ref{eq:R_tot_fm}). This can be understood as the effect of the reduction of spin relaxation events.

In summary, we have shown that the Hanle spin precession-induced magnetoresistance for a chiral system on a semiconductor is inverted compared to the typical ferromagnetic Hanle case. With a chiral element as injector and detector, the build-up of a spin accumulation reduces the total resistance of the chiral element/semiconductor. This result is universal for any chiral system as they were obtained using the most general spin-dependent transport model for a chiral element. Additionally, we have explicitly included spin relaxation processes in the semiconductor and verified that our results remain true with them included. 

This work is supported by the Zernike Institute for Advanced Materials (ZIAM) and the Spinoza prize awarded to Professor B.J. van Wees by the Nederlandse Organisatie voor Wetenschappelijk Onderzoek (NWO) in 2016, as well as by the European Research Council (ERC) under the European Union's 2DMAGSPIN (Grant Agreement No. 101053054). We also thank T. Banerjee and M. Cosset-Chéneau for valuable discussions and comments on the manuscript.

\section*{AUTHOR DECLARATIONS}
\subsection*{Conflict of Interest}
The authors have no conflicts to disclose.

\subsection*{Author Contributions}
\textbf{Sytze H. Tirion}: Conceptualization (equal); Formal analysis (equal); Investigation (equal); Writing - original draft; (equal); Writing - review $\&$ editing (equal). \textbf{Bart J. van Wees}: Conceptualization (equal); Formal analysis (equal); Investigation (equal); Writing - original draft; (equal); Writing - review $\&$ editing (equal); Supervision (lead).

\section*{REFERENCES}
\nocite{*}
\bibliography{aipsamp}

\providecommand{\noopsort}[1]{}\providecommand{\singleletter}[1]{#1}%
\begin{thebibliography}{38}%
\makeatletter
\providecommand \@ifxundefined [1]{%
 \@ifx{#1\undefined}
}%
\providecommand \@ifnum [1]{%
 \ifnum #1\expandafter \@firstoftwo
 \else \expandafter \@secondoftwo
 \fi
}%
\providecommand \@ifx [1]{%
 \ifx #1\expandafter \@firstoftwo
 \else \expandafter \@secondoftwo
 \fi
}%
\providecommand \natexlab [1]{#1}%
\providecommand \enquote  [1]{``#1''}%
\providecommand \bibnamefont  [1]{#1}%
\providecommand \bibfnamefont [1]{#1}%
\providecommand \citenamefont [1]{#1}%
\providecommand \href@noop [0]{\@secondoftwo}%
\providecommand \href [0]{\begingroup \@sanitize@url \@href}%
\providecommand \@href[1]{\@@startlink{#1}\@@href}%
\providecommand \@@href[1]{\endgroup#1\@@endlink}%
\providecommand \@sanitize@url [0]{\catcode `\\12\catcode `\$12\catcode `\&12\catcode `\#12\catcode `\^12\catcode `\_12\catcode `\%12\relax}%
\providecommand \@@startlink[1]{}%
\providecommand \@@endlink[0]{}%
\providecommand \url  [0]{\begingroup\@sanitize@url \@url }%
\providecommand \@url [1]{\endgroup\@href {#1}{\urlprefix }}%
\providecommand \urlprefix  [0]{URL }%
\providecommand \Eprint [0]{\href }%
\providecommand \doibase [0]{http://dx.doi.org/}%
\providecommand \selectlanguage [0]{\@gobble}%
\providecommand \bibinfo  [0]{\@secondoftwo}%
\providecommand \bibfield  [0]{\@secondoftwo}%
\providecommand \translation [1]{[#1]}%
\providecommand \BibitemOpen [0]{}%
\providecommand \bibitemStop [0]{}%
\providecommand \bibitemNoStop [0]{.\EOS\space}%
\providecommand \EOS [0]{\spacefactor3000\relax}%
\providecommand \BibitemShut  [1]{\csname bibitem#1\endcsname}%
\let\auto@bib@innerbib\@empty
\bibitem [{\citenamefont {Spiesser}\ \emph {et~al.}(2019)\citenamefont {Spiesser}, \citenamefont {Fujita}, \citenamefont {Saito}, \citenamefont {Yamada}, \citenamefont {Hamaya}, \citenamefont {Yuasa},\ and\ \citenamefont {Jansen}}]{Spiesser2019}%
  \BibitemOpen
  \bibfield  {author} {\bibinfo {author} {\bibfnamefont {A.}~\bibnamefont {Spiesser}}, \bibinfo {author} {\bibfnamefont {Y.}~\bibnamefont {Fujita}}, \bibinfo {author} {\bibfnamefont {H.}~\bibnamefont {Saito}}, \bibinfo {author} {\bibfnamefont {S.}~\bibnamefont {Yamada}}, \bibinfo {author} {\bibfnamefont {K.}~\bibnamefont {Hamaya}}, \bibinfo {author} {\bibfnamefont {S.}~\bibnamefont {Yuasa}}, \ and\ \bibinfo {author} {\bibfnamefont {R.}~\bibnamefont {Jansen}},\ }\bibfield  {title} {\enquote {\bibinfo {title} {{Hanle spin precession in a two-terminal lateral spin valve}},}\ }\href@noop {} {\bibfield  {journal} {\bibinfo  {journal} {Applied Physics Letters}\ }\textbf {\bibinfo {volume} {114}},\ \bibinfo {pages} {242401} (\bibinfo {year} {2019})}\BibitemShut {NoStop}%
\bibitem [{\citenamefont {Dash}\ \emph {et~al.}(2009)\citenamefont {Dash}, \citenamefont {Sharma}, \citenamefont {Patel}, \citenamefont {{De Jong}},\ and\ \citenamefont {Jansen}}]{Dash2009}%
  \BibitemOpen
  \bibfield  {author} {\bibinfo {author} {\bibfnamefont {S.~P.}\ \bibnamefont {Dash}}, \bibinfo {author} {\bibfnamefont {S.}~\bibnamefont {Sharma}}, \bibinfo {author} {\bibfnamefont {R.~S.}\ \bibnamefont {Patel}}, \bibinfo {author} {\bibfnamefont {M.~P.}\ \bibnamefont {{De Jong}}}, \ and\ \bibinfo {author} {\bibfnamefont {R.}~\bibnamefont {Jansen}},\ }\bibfield  {title} {\enquote {\bibinfo {title} {{Electrical creation of spin polarization in silicon at room temperature}},}\ }\href@noop {} {\bibfield  {journal} {\bibinfo  {journal} {Nature}\ }\textbf {\bibinfo {volume} {462}},\ \bibinfo {pages} {491--494} (\bibinfo {year} {2009})}\BibitemShut {NoStop}%
\bibitem [{\citenamefont {Lou}\ \emph {et~al.}(2006)\citenamefont {Lou}, \citenamefont {Adelmann}, \citenamefont {Furis}, \citenamefont {Crooker}, \citenamefont {Palmstr{\o}m},\ and\ \citenamefont {Crowell}}]{Lou2006}%
  \BibitemOpen
  \bibfield  {author} {\bibinfo {author} {\bibfnamefont {X.}~\bibnamefont {Lou}}, \bibinfo {author} {\bibfnamefont {C.}~\bibnamefont {Adelmann}}, \bibinfo {author} {\bibfnamefont {M.}~\bibnamefont {Furis}}, \bibinfo {author} {\bibfnamefont {S.~A.}\ \bibnamefont {Crooker}}, \bibinfo {author} {\bibfnamefont {C.~J.}\ \bibnamefont {Palmstr{\o}m}}, \ and\ \bibinfo {author} {\bibfnamefont {P.~A.}\ \bibnamefont {Crowell}},\ }\bibfield  {title} {\enquote {\bibinfo {title} {{Electrical Detection of Spin Accumulation at a Ferromagnet-Semiconductor Interface}},}\ }\href@noop {} {\bibfield  {journal} {\bibinfo  {journal} {Physical Review Letters}\ }\textbf {\bibinfo {volume} {96}},\ \bibinfo {pages} {176603} (\bibinfo {year} {2006})}\BibitemShut {NoStop}%
\bibitem [{\citenamefont {Kamerbeek}\ \emph {et~al.}(2015)\citenamefont {Kamerbeek}, \citenamefont {H{\"{o}}gl}, \citenamefont {Fabian},\ and\ \citenamefont {Banerjee}}]{Kamerbeek2015}%
  \BibitemOpen
  \bibfield  {author} {\bibinfo {author} {\bibfnamefont {A.~M.}\ \bibnamefont {Kamerbeek}}, \bibinfo {author} {\bibfnamefont {P.}~\bibnamefont {H{\"{o}}gl}}, \bibinfo {author} {\bibfnamefont {J.}~\bibnamefont {Fabian}}, \ and\ \bibinfo {author} {\bibfnamefont {T.}~\bibnamefont {Banerjee}},\ }\bibfield  {title} {\enquote {\bibinfo {title} {{Electric Field Control of Spin Lifetimes in Nb-SrTiO3 by Spin-Orbit Fields}},}\ }\href@noop {} {\bibfield  {journal} {\bibinfo  {journal} {Physical Review Letters}\ }\textbf {\bibinfo {volume} {115}},\ \bibinfo {pages} {136601} (\bibinfo {year} {2015})}\BibitemShut {NoStop}%
\bibitem [{\citenamefont {Tombros}\ \emph {et~al.}(2007)\citenamefont {Tombros}, \citenamefont {Jozsa}, \citenamefont {Popinciuc}, \citenamefont {Jonkman},\ and\ \citenamefont {van Wees}}]{Tombros2007}%
  \BibitemOpen
  \bibfield  {author} {\bibinfo {author} {\bibfnamefont {N.}~\bibnamefont {Tombros}}, \bibinfo {author} {\bibfnamefont {C.}~\bibnamefont {Jozsa}}, \bibinfo {author} {\bibfnamefont {M.}~\bibnamefont {Popinciuc}}, \bibinfo {author} {\bibfnamefont {H.~T.}\ \bibnamefont {Jonkman}}, \ and\ \bibinfo {author} {\bibfnamefont {B.~J.}\ \bibnamefont {van Wees}},\ }\bibfield  {title} {\enquote {\bibinfo {title} {{Electronic spin transport and spin precession in single graphene layers at room temperature}},}\ }\href@noop {} {\bibfield  {journal} {\bibinfo  {journal} {Nature}\ }\textbf {\bibinfo {volume} {448}},\ \bibinfo {pages} {571--574} (\bibinfo {year} {2007})}\BibitemShut {NoStop}%
\bibitem [{\citenamefont {Hu}\ and\ \citenamefont {Xiang}(2020)}]{Hu2020}%
  \BibitemOpen
  \bibfield  {author} {\bibinfo {author} {\bibfnamefont {G.}~\bibnamefont {Hu}}\ and\ \bibinfo {author} {\bibfnamefont {B.}~\bibnamefont {Xiang}},\ }\bibfield  {title} {\enquote {\bibinfo {title} {{Recent Advances in Two-Dimensional Spintronics}},}\ }\href@noop {} {\bibfield  {journal} {\bibinfo  {journal} {Nanoscale Research Letters}\ }\textbf {\bibinfo {volume} {15}},\ \bibinfo {pages} {226} (\bibinfo {year} {2020})}\BibitemShut {NoStop}%
\bibitem [{\citenamefont {V{\'{e}}lez}\ \emph {et~al.}(2016)\citenamefont {V{\'{e}}lez}, \citenamefont {Golovach}, \citenamefont {Bedoya-Pinto}, \citenamefont {Isasa}, \citenamefont {Sagasta}, \citenamefont {Abadia}, \citenamefont {Rogero}, \citenamefont {Hueso}, \citenamefont {Bergeret},\ and\ \citenamefont {Casanova}}]{Velez2016}%
  \BibitemOpen
  \bibfield  {author} {\bibinfo {author} {\bibfnamefont {S.}~\bibnamefont {V{\'{e}}lez}}, \bibinfo {author} {\bibfnamefont {V.~N.}\ \bibnamefont {Golovach}}, \bibinfo {author} {\bibfnamefont {A.}~\bibnamefont {Bedoya-Pinto}}, \bibinfo {author} {\bibfnamefont {M.}~\bibnamefont {Isasa}}, \bibinfo {author} {\bibfnamefont {E.}~\bibnamefont {Sagasta}}, \bibinfo {author} {\bibfnamefont {M.}~\bibnamefont {Abadia}}, \bibinfo {author} {\bibfnamefont {C.}~\bibnamefont {Rogero}}, \bibinfo {author} {\bibfnamefont {L.~E.}\ \bibnamefont {Hueso}}, \bibinfo {author} {\bibfnamefont {F.~S.}\ \bibnamefont {Bergeret}}, \ and\ \bibinfo {author} {\bibfnamefont {F.}~\bibnamefont {Casanova}},\ }\bibfield  {title} {\enquote {\bibinfo {title} {{Hanle Magnetoresistance in Thin Metal Films with Strong Spin-Orbit Coupling}},}\ }\href@noop {} {\bibfield  {journal} {\bibinfo  {journal} {Physical Review Letters}\ }\textbf {\bibinfo {volume} {116}},\ \bibinfo {pages} {016603} (\bibinfo {year} {2016})}\BibitemShut {NoStop}%
\bibitem [{\citenamefont {Jedema}\ \emph {et~al.}(2002)\citenamefont {Jedema}, \citenamefont {Heersche}, \citenamefont {Filip}, \citenamefont {Baselmans},\ and\ \citenamefont {van Wees}}]{Jedema2002}%
  \BibitemOpen
  \bibfield  {author} {\bibinfo {author} {\bibfnamefont {F.}~\bibnamefont {Jedema}}, \bibinfo {author} {\bibfnamefont {H.}~\bibnamefont {Heersche}}, \bibinfo {author} {\bibfnamefont {A.}~\bibnamefont {Filip}}, \bibinfo {author} {\bibfnamefont {J.}~\bibnamefont {Baselmans}}, \ and\ \bibinfo {author} {\bibfnamefont {B.~J.}\ \bibnamefont {van Wees}},\ }\bibfield  {title} {\enquote {\bibinfo {title} {{Electrical detection of spin precession in a metallic mesoscopic spin valve}},}\ }\href@noop {} {\bibfield  {journal} {\bibinfo  {journal} {Nature}\ }\textbf {\bibinfo {volume} {416}},\ \bibinfo {pages} {713--716} (\bibinfo {year} {2002})}\BibitemShut {NoStop}%
\bibitem [{\citenamefont {Yang}, \citenamefont {{Van Der Wal}},\ and\ \citenamefont {{Van Wees}}(2020)}]{Yang2020a}%
  \BibitemOpen
  \bibfield  {author} {\bibinfo {author} {\bibfnamefont {X.}~\bibnamefont {Yang}}, \bibinfo {author} {\bibfnamefont {C.~H.}\ \bibnamefont {{Van Der Wal}}}, \ and\ \bibinfo {author} {\bibfnamefont {B.~J.}\ \bibnamefont {{Van Wees}}},\ }\bibfield  {title} {\enquote {\bibinfo {title} {{Detecting Chirality in Two-Terminal Electronic Nanodevices}},}\ }\href@noop {} {\bibfield  {journal} {\bibinfo  {journal} {Nano Letters}\ }\textbf {\bibinfo {volume} {20}},\ \bibinfo {pages} {6148--6154} (\bibinfo {year} {2020})}\BibitemShut {NoStop}%
\bibitem [{\citenamefont {Yang}\ and\ \citenamefont {van Wees}(2021)}]{Yang2021}%
  \BibitemOpen
  \bibfield  {author} {\bibinfo {author} {\bibfnamefont {X.}~\bibnamefont {Yang}}\ and\ \bibinfo {author} {\bibfnamefont {B.~J.}\ \bibnamefont {van Wees}},\ }\bibfield  {title} {\enquote {\bibinfo {title} {{Linear-response magnetoresistance effects in chiral systems}},}\ }\href@noop {} {\bibfield  {journal} {\bibinfo  {journal} {Physical Review B}\ }\textbf {\bibinfo {volume} {104}},\ \bibinfo {pages} {155420} (\bibinfo {year} {2021})}\BibitemShut {NoStop}%
\bibitem [{\citenamefont {Sun}\ \emph {et~al.}(2024)\citenamefont {Sun}, \citenamefont {Park}, \citenamefont {Comstock}, \citenamefont {McConnell}, \citenamefont {Chen}, \citenamefont {Zhang}, \citenamefont {Beratan}, \citenamefont {You}, \citenamefont {Hoffmann}, \citenamefont {Yu}, \citenamefont {Diao},\ and\ \citenamefont {Sun}}]{Sun2024}%
  \BibitemOpen
  \bibfield  {author} {\bibinfo {author} {\bibfnamefont {R.}~\bibnamefont {Sun}}, \bibinfo {author} {\bibfnamefont {K.~S.}\ \bibnamefont {Park}}, \bibinfo {author} {\bibfnamefont {A.~H.}\ \bibnamefont {Comstock}}, \bibinfo {author} {\bibfnamefont {A.}~\bibnamefont {McConnell}}, \bibinfo {author} {\bibfnamefont {Y.-c.}\ \bibnamefont {Chen}}, \bibinfo {author} {\bibfnamefont {P.}~\bibnamefont {Zhang}}, \bibinfo {author} {\bibfnamefont {D.}~\bibnamefont {Beratan}}, \bibinfo {author} {\bibfnamefont {W.}~\bibnamefont {You}}, \bibinfo {author} {\bibfnamefont {A.}~\bibnamefont {Hoffmann}}, \bibinfo {author} {\bibfnamefont {Z.-g.}\ \bibnamefont {Yu}}, \bibinfo {author} {\bibfnamefont {Y.}~\bibnamefont {Diao}}, \ and\ \bibinfo {author} {\bibfnamefont {D.}~\bibnamefont {Sun}},\ }\bibfield  {title} {\enquote {\bibinfo {title} {{Inverse chirality-induced spin selectivity effect in chiral assemblies of $\pi$-conjugated polymers}},}\ }\href@noop {} {\bibfield  {journal} {\bibinfo  {journal} {Nature Materials}\ }\textbf
  {\bibinfo {volume} {23}},\ \bibinfo {pages} {782--789} (\bibinfo {year} {2024})}\BibitemShut {NoStop}%
\bibitem [{\citenamefont {Liu}\ \emph {et~al.}(2024)\citenamefont {Liu}, \citenamefont {Adhikari}, \citenamefont {Wang}, \citenamefont {Jiang}, \citenamefont {Hua}, \citenamefont {Liu}, \citenamefont {Schlottmann}, \citenamefont {Gao}, \citenamefont {Weiss}, \citenamefont {Yan}, \citenamefont {Zhao},\ and\ \citenamefont {Xiong}}]{Liu2024}%
  \BibitemOpen
  \bibfield  {author} {\bibinfo {author} {\bibfnamefont {T.}~\bibnamefont {Liu}}, \bibinfo {author} {\bibfnamefont {Y.}~\bibnamefont {Adhikari}}, \bibinfo {author} {\bibfnamefont {H.}~\bibnamefont {Wang}}, \bibinfo {author} {\bibfnamefont {Y.}~\bibnamefont {Jiang}}, \bibinfo {author} {\bibfnamefont {Z.}~\bibnamefont {Hua}}, \bibinfo {author} {\bibfnamefont {H.}~\bibnamefont {Liu}}, \bibinfo {author} {\bibfnamefont {P.}~\bibnamefont {Schlottmann}}, \bibinfo {author} {\bibfnamefont {H.}~\bibnamefont {Gao}}, \bibinfo {author} {\bibfnamefont {P.~S.}\ \bibnamefont {Weiss}}, \bibinfo {author} {\bibfnamefont {B.}~\bibnamefont {Yan}}, \bibinfo {author} {\bibfnamefont {J.}~\bibnamefont {Zhao}}, \ and\ \bibinfo {author} {\bibfnamefont {P.}~\bibnamefont {Xiong}},\ }\bibfield  {title} {\enquote {\bibinfo {title} {{Chirality‐Induced Magnet‐Free Spin Generation in a Semiconductor}},}\ }\href@noop {} {\bibfield  {journal} {\bibinfo  {journal} {Advanced Materials}\ }\textbf {\bibinfo {volume} {36}},\ \bibinfo {pages}
  {2406347} (\bibinfo {year} {2024})}\BibitemShut {NoStop}%
\bibitem [{\citenamefont {Bloom}\ \emph {et~al.}(2024)\citenamefont {Bloom}, \citenamefont {Paltiel}, \citenamefont {Naaman},\ and\ \citenamefont {Waldeck}}]{Bloom2024}%
  \BibitemOpen
  \bibfield  {author} {\bibinfo {author} {\bibfnamefont {B.~P.}\ \bibnamefont {Bloom}}, \bibinfo {author} {\bibfnamefont {Y.}~\bibnamefont {Paltiel}}, \bibinfo {author} {\bibfnamefont {R.}~\bibnamefont {Naaman}}, \ and\ \bibinfo {author} {\bibfnamefont {D.~H.}\ \bibnamefont {Waldeck}},\ }\bibfield  {title} {\enquote {\bibinfo {title} {{Chiral Induced Spin Selectivity}},}\ }\href@noop {} {\bibfield  {journal} {\bibinfo  {journal} {Chemical Reviews}\ }\textbf {\bibinfo {volume} {124}},\ \bibinfo {pages} {1950--1991} (\bibinfo {year} {2024})}\BibitemShut {NoStop}%
\bibitem [{\citenamefont {Firouzeh}\ \emph {et~al.}(2024)\citenamefont {Firouzeh}, \citenamefont {Hossain}, \citenamefont {Cuerva}, \citenamefont {{{\'{A}}lvarez de Cienfuegos}},\ and\ \citenamefont {Pramanik}}]{Firouzeh2024}%
  \BibitemOpen
  \bibfield  {author} {\bibinfo {author} {\bibfnamefont {S.}~\bibnamefont {Firouzeh}}, \bibinfo {author} {\bibfnamefont {M.~A.}\ \bibnamefont {Hossain}}, \bibinfo {author} {\bibfnamefont {J.~M.}\ \bibnamefont {Cuerva}}, \bibinfo {author} {\bibfnamefont {L.}~\bibnamefont {{{\'{A}}lvarez de Cienfuegos}}}, \ and\ \bibinfo {author} {\bibfnamefont {S.}~\bibnamefont {Pramanik}},\ }\bibfield  {title} {\enquote {\bibinfo {title} {{Chirality-Induced Spin Selectivity in Composite Materials: A Device Perspective}},}\ }\href@noop {} {\bibfield  {journal} {\bibinfo  {journal} {Accounts of Chemical Research}\ }\textbf {\bibinfo {volume} {57}},\ \bibinfo {pages} {1478--1487} (\bibinfo {year} {2024})}\BibitemShut {NoStop}%
\bibitem [{\citenamefont {Evers}\ \emph {et~al.}(2022)\citenamefont {Evers}, \citenamefont {Aharony}, \citenamefont {Bar-Gill}, \citenamefont {Entin-Wohlman}, \citenamefont {Hedeg{\aa}rd}, \citenamefont {Hod}, \citenamefont {Jelinek}, \citenamefont {Kamieniarz}, \citenamefont {Lemeshko}, \citenamefont {Michaeli}, \citenamefont {Mujica}, \citenamefont {Naaman}, \citenamefont {Paltiel}, \citenamefont {Refaely-Abramson}, \citenamefont {Tal}, \citenamefont {Thijssen}, \citenamefont {Thoss}, \citenamefont {van Ruitenbeek}, \citenamefont {Venkataraman}, \citenamefont {Waldeck}, \citenamefont {Yan},\ and\ \citenamefont {Kronik}}]{Evers2022}%
  \BibitemOpen
  \bibfield  {author} {\bibinfo {author} {\bibfnamefont {F.}~\bibnamefont {Evers}}, \bibinfo {author} {\bibfnamefont {A.}~\bibnamefont {Aharony}}, \bibinfo {author} {\bibfnamefont {N.}~\bibnamefont {Bar-Gill}}, \bibinfo {author} {\bibfnamefont {O.}~\bibnamefont {Entin-Wohlman}}, \bibinfo {author} {\bibfnamefont {P.}~\bibnamefont {Hedeg{\aa}rd}}, \bibinfo {author} {\bibfnamefont {O.}~\bibnamefont {Hod}}, \bibinfo {author} {\bibfnamefont {P.}~\bibnamefont {Jelinek}}, \bibinfo {author} {\bibfnamefont {G.}~\bibnamefont {Kamieniarz}}, \bibinfo {author} {\bibfnamefont {M.}~\bibnamefont {Lemeshko}}, \bibinfo {author} {\bibfnamefont {K.}~\bibnamefont {Michaeli}}, \bibinfo {author} {\bibfnamefont {V.}~\bibnamefont {Mujica}}, \bibinfo {author} {\bibfnamefont {R.}~\bibnamefont {Naaman}}, \bibinfo {author} {\bibfnamefont {Y.}~\bibnamefont {Paltiel}}, \bibinfo {author} {\bibfnamefont {S.}~\bibnamefont {Refaely-Abramson}}, \bibinfo {author} {\bibfnamefont {O.}~\bibnamefont {Tal}}, \bibinfo {author} {\bibfnamefont
  {J.}~\bibnamefont {Thijssen}}, \bibinfo {author} {\bibfnamefont {M.}~\bibnamefont {Thoss}}, \bibinfo {author} {\bibfnamefont {J.~M.}\ \bibnamefont {van Ruitenbeek}}, \bibinfo {author} {\bibfnamefont {L.}~\bibnamefont {Venkataraman}}, \bibinfo {author} {\bibfnamefont {D.~H.}\ \bibnamefont {Waldeck}}, \bibinfo {author} {\bibfnamefont {B.}~\bibnamefont {Yan}}, \ and\ \bibinfo {author} {\bibfnamefont {L.}~\bibnamefont {Kronik}},\ }\bibfield  {title} {\enquote {\bibinfo {title} {{Theory of Chirality Induced Spin Selectivity: Progress and Challenges}},}\ }\href@noop {} {\bibfield  {journal} {\bibinfo  {journal} {Advanced Materials}\ }\textbf {\bibinfo {volume} {34}},\ \bibinfo {pages} {2106629} (\bibinfo {year} {2022})}\BibitemShut {NoStop}%
\bibitem [{\citenamefont {Mishra}\ \emph {et~al.}(2020)\citenamefont {Mishra}, \citenamefont {Mondal}, \citenamefont {Smolinsky}, \citenamefont {Naaman}, \citenamefont {Maeda}, \citenamefont {Nishimura}, \citenamefont {Taniguchi}, \citenamefont {Yoshida}, \citenamefont {Takayama},\ and\ \citenamefont {Yashima}}]{Mishra2020}%
  \BibitemOpen
  \bibfield  {author} {\bibinfo {author} {\bibfnamefont {S.~M.}\ \bibnamefont {Mishra}}, \bibinfo {author} {\bibfnamefont {A.~K.}\ \bibnamefont {Mondal}}, \bibinfo {author} {\bibfnamefont {E.~Z.~B.}\ \bibnamefont {Smolinsky}}, \bibinfo {author} {\bibfnamefont {R.}~\bibnamefont {Naaman}}, \bibinfo {author} {\bibfnamefont {K.}~\bibnamefont {Maeda}}, \bibinfo {author} {\bibfnamefont {T.}~\bibnamefont {Nishimura}}, \bibinfo {author} {\bibfnamefont {T.}~\bibnamefont {Taniguchi}}, \bibinfo {author} {\bibfnamefont {T.}~\bibnamefont {Yoshida}}, \bibinfo {author} {\bibfnamefont {K.}~\bibnamefont {Takayama}}, \ and\ \bibinfo {author} {\bibfnamefont {E.}~\bibnamefont {Yashima}},\ }\bibfield  {title} {\enquote {\bibinfo {title} {{Spin Filtering Along Chiral Polymers}},}\ }\href@noop {} {\bibfield  {journal} {\bibinfo  {journal} {Angewandte Chemie - International Edition}\ }\textbf {\bibinfo {volume} {59}},\ \bibinfo {pages} {14671--14676} (\bibinfo {year} {2020})}\BibitemShut {NoStop}%
\bibitem [{\citenamefont {Liu}\ \emph {et~al.}(2020)\citenamefont {Liu}, \citenamefont {Wang}, \citenamefont {Wang}, \citenamefont {Shi}, \citenamefont {Gao}, \citenamefont {Feng}, \citenamefont {Deng}, \citenamefont {Hu}, \citenamefont {Lochner}, \citenamefont {Schlottmann}, \citenamefont {{Von Moln{\'{a}}r}}, \citenamefont {Li}, \citenamefont {Zhao},\ and\ \citenamefont {Xiong}}]{Liu2020}%
  \BibitemOpen
  \bibfield  {author} {\bibinfo {author} {\bibfnamefont {T.}~\bibnamefont {Liu}}, \bibinfo {author} {\bibfnamefont {X.}~\bibnamefont {Wang}}, \bibinfo {author} {\bibfnamefont {H.}~\bibnamefont {Wang}}, \bibinfo {author} {\bibfnamefont {G.}~\bibnamefont {Shi}}, \bibinfo {author} {\bibfnamefont {F.}~\bibnamefont {Gao}}, \bibinfo {author} {\bibfnamefont {H.}~\bibnamefont {Feng}}, \bibinfo {author} {\bibfnamefont {H.}~\bibnamefont {Deng}}, \bibinfo {author} {\bibfnamefont {L.}~\bibnamefont {Hu}}, \bibinfo {author} {\bibfnamefont {E.}~\bibnamefont {Lochner}}, \bibinfo {author} {\bibfnamefont {P.}~\bibnamefont {Schlottmann}}, \bibinfo {author} {\bibfnamefont {S.}~\bibnamefont {{Von Moln{\'{a}}r}}}, \bibinfo {author} {\bibfnamefont {Y.}~\bibnamefont {Li}}, \bibinfo {author} {\bibfnamefont {J.}~\bibnamefont {Zhao}}, \ and\ \bibinfo {author} {\bibfnamefont {P.}~\bibnamefont {Xiong}},\ }\bibfield  {title} {\enquote {\bibinfo {title} {{Linear and nonlinear two-terminal spin-valve effect from chirality-induced spin
  selectivity}},}\ }\href@noop {} {\bibfield  {journal} {\bibinfo  {journal} {ACS Nano}\ }\textbf {\bibinfo {volume} {14}},\ \bibinfo {pages} {15983--15991} (\bibinfo {year} {2020})}\BibitemShut {NoStop}%
\bibitem [{\citenamefont {Mondal}\ \emph {et~al.}(2021)\citenamefont {Mondal}, \citenamefont {Preuss}, \citenamefont {{\'{S}}l\c{e}czkowski}, \citenamefont {Das}, \citenamefont {Vantomme}, \citenamefont {Meijer},\ and\ \citenamefont {Naaman}}]{Mondal2021}%
  \BibitemOpen
  \bibfield  {author} {\bibinfo {author} {\bibfnamefont {A.~K.}\ \bibnamefont {Mondal}}, \bibinfo {author} {\bibfnamefont {M.~D.}\ \bibnamefont {Preuss}}, \bibinfo {author} {\bibfnamefont {M.~L.}\ \bibnamefont {{\'{S}}l\c{e}czkowski}}, \bibinfo {author} {\bibfnamefont {T.~K.}\ \bibnamefont {Das}}, \bibinfo {author} {\bibfnamefont {G.}~\bibnamefont {Vantomme}}, \bibinfo {author} {\bibfnamefont {E.~W.}\ \bibnamefont {Meijer}}, \ and\ \bibinfo {author} {\bibfnamefont {R.}~\bibnamefont {Naaman}},\ }\bibfield  {title} {\enquote {\bibinfo {title} {{Spin Filtering in Supramolecular Polymers Assembled from Achiral Monomers Mediated by Chiral Solvents}},}\ }\href@noop {} {\bibfield  {journal} {\bibinfo  {journal} {Journal of the American Chemical Society}\ }\textbf {\bibinfo {volume} {143}},\ \bibinfo {pages} {7189--7195} (\bibinfo {year} {2021})}\BibitemShut {NoStop}%
\bibitem [{\citenamefont {Safari}\ \emph {et~al.}(2023)\citenamefont {Safari}, \citenamefont {Matthes}, \citenamefont {Schneider}, \citenamefont {Ernst},\ and\ \citenamefont {B{\"{u}}rgler}}]{Safari2023}%
  \BibitemOpen
  \bibfield  {author} {\bibinfo {author} {\bibfnamefont {M.~R.}\ \bibnamefont {Safari}}, \bibinfo {author} {\bibfnamefont {F.}~\bibnamefont {Matthes}}, \bibinfo {author} {\bibfnamefont {C.~M.}\ \bibnamefont {Schneider}}, \bibinfo {author} {\bibfnamefont {K.-H.}\ \bibnamefont {Ernst}}, \ and\ \bibinfo {author} {\bibfnamefont {D.~E.}\ \bibnamefont {B{\"{u}}rgler}},\ }\bibfield  {title} {\enquote {\bibinfo {title} {{Spin‐Selective Electron Transport Through Single Chiral Molecules}},}\ }\href@noop {} {\bibfield  {journal} {\bibinfo  {journal} {Small}\ }\textbf {\bibinfo {volume} {20}},\ \bibinfo {pages} {2308233} (\bibinfo {year} {2023})}\BibitemShut {NoStop}%
\bibitem [{\citenamefont {Firouzeh}\ \emph {et~al.}(2023)\citenamefont {Firouzeh}, \citenamefont {Illescas-Lopez}, \citenamefont {Hossain}, \citenamefont {Cuerva}, \citenamefont {{{\'{A}}lvarez de Cienfuegos}},\ and\ \citenamefont {Pramanik}}]{Firouzeh2023}%
  \BibitemOpen
  \bibfield  {author} {\bibinfo {author} {\bibfnamefont {S.}~\bibnamefont {Firouzeh}}, \bibinfo {author} {\bibfnamefont {S.}~\bibnamefont {Illescas-Lopez}}, \bibinfo {author} {\bibfnamefont {M.~A.}\ \bibnamefont {Hossain}}, \bibinfo {author} {\bibfnamefont {J.~M.}\ \bibnamefont {Cuerva}}, \bibinfo {author} {\bibfnamefont {L.}~\bibnamefont {{{\'{A}}lvarez de Cienfuegos}}}, \ and\ \bibinfo {author} {\bibfnamefont {S.}~\bibnamefont {Pramanik}},\ }\bibfield  {title} {\enquote {\bibinfo {title} {{Chirality-Induced Spin Selectivity in Supramolecular Chirally Functionalized Graphene}},}\ }\href@noop {} {\bibfield  {journal} {\bibinfo  {journal} {ACS Nano}\ }\textbf {\bibinfo {volume} {17}},\ \bibinfo {pages} {20424--20433} (\bibinfo {year} {2023})}\BibitemShut {NoStop}%
\bibitem [{\citenamefont {Singh}\ \emph {et~al.}(2024)\citenamefont {Singh}, \citenamefont {Martin}, \citenamefont {Talamo}, \citenamefont {Houssin}, \citenamefont {Vanthuyne}, \citenamefont {Avarvari},\ and\ \citenamefont {Tal}}]{Singh2024}%
  \BibitemOpen
  \bibfield  {author} {\bibinfo {author} {\bibfnamefont {A.}~\bibnamefont {Singh}}, \bibinfo {author} {\bibfnamefont {K.}~\bibnamefont {Martin}}, \bibinfo {author} {\bibfnamefont {M.~M.}\ \bibnamefont {Talamo}}, \bibinfo {author} {\bibfnamefont {A.}~\bibnamefont {Houssin}}, \bibinfo {author} {\bibfnamefont {N.}~\bibnamefont {Vanthuyne}}, \bibinfo {author} {\bibfnamefont {N.}~\bibnamefont {Avarvari}}, \ and\ \bibinfo {author} {\bibfnamefont {O.}~\bibnamefont {Tal}},\ }\href@noop {} {\enquote {\bibinfo {title} {Single-molecule junctions map the interplay between electrons and chirality},}\ } (\bibinfo {year} {2024}),\ \bibinfo {note} {preprint, doi: 10.48550/arXiv.2408.12258V1 (Accessed September 1, 2023)}\BibitemShut {NoStop}%
\bibitem [{\citenamefont {Lu}\ \emph {et~al.}(2019)\citenamefont {Lu}, \citenamefont {Wang}, \citenamefont {Xiao}, \citenamefont {Pan}, \citenamefont {Chen}, \citenamefont {Brunecky}, \citenamefont {Berry}, \citenamefont {Zhu}, \citenamefont {Beard},\ and\ \citenamefont {Vardeny}}]{Lu2019}%
  \BibitemOpen
  \bibfield  {author} {\bibinfo {author} {\bibfnamefont {H.}~\bibnamefont {Lu}}, \bibinfo {author} {\bibfnamefont {J.}~\bibnamefont {Wang}}, \bibinfo {author} {\bibfnamefont {C.}~\bibnamefont {Xiao}}, \bibinfo {author} {\bibfnamefont {X.}~\bibnamefont {Pan}}, \bibinfo {author} {\bibfnamefont {X.}~\bibnamefont {Chen}}, \bibinfo {author} {\bibfnamefont {R.}~\bibnamefont {Brunecky}}, \bibinfo {author} {\bibfnamefont {J.~J.}\ \bibnamefont {Berry}}, \bibinfo {author} {\bibfnamefont {K.}~\bibnamefont {Zhu}}, \bibinfo {author} {\bibfnamefont {M.~C.}\ \bibnamefont {Beard}}, \ and\ \bibinfo {author} {\bibfnamefont {Z.~V.}\ \bibnamefont {Vardeny}},\ }\bibfield  {title} {\enquote {\bibinfo {title} {{Spin-dependent charge transport through 2D chiral hybrid lead-iodide perovskites}},}\ }\href@noop {} {\bibfield  {journal} {\bibinfo  {journal} {Science Advances}\ }\textbf {\bibinfo {volume} {5}},\ \bibinfo {pages} {1--8} (\bibinfo {year} {2019})}\BibitemShut {NoStop}%
\bibitem [{\citenamefont {Inui}\ \emph {et~al.}(2020)\citenamefont {Inui}, \citenamefont {Aoki}, \citenamefont {Nishiue}, \citenamefont {Shiota}, \citenamefont {Kousaka}, \citenamefont {Shishido}, \citenamefont {Hirobe}, \citenamefont {Suda}, \citenamefont {Ohe}, \citenamefont {Kishine}, \citenamefont {Yamamoto},\ and\ \citenamefont {Togawa}}]{Inui2020}%
  \BibitemOpen
  \bibfield  {author} {\bibinfo {author} {\bibfnamefont {A.}~\bibnamefont {Inui}}, \bibinfo {author} {\bibfnamefont {R.}~\bibnamefont {Aoki}}, \bibinfo {author} {\bibfnamefont {Y.}~\bibnamefont {Nishiue}}, \bibinfo {author} {\bibfnamefont {K.}~\bibnamefont {Shiota}}, \bibinfo {author} {\bibfnamefont {Y.}~\bibnamefont {Kousaka}}, \bibinfo {author} {\bibfnamefont {H.}~\bibnamefont {Shishido}}, \bibinfo {author} {\bibfnamefont {D.}~\bibnamefont {Hirobe}}, \bibinfo {author} {\bibfnamefont {M.}~\bibnamefont {Suda}}, \bibinfo {author} {\bibfnamefont {J.~I.}\ \bibnamefont {Ohe}}, \bibinfo {author} {\bibfnamefont {J.~I.}\ \bibnamefont {Kishine}}, \bibinfo {author} {\bibfnamefont {H.~M.}\ \bibnamefont {Yamamoto}}, \ and\ \bibinfo {author} {\bibfnamefont {Y.}~\bibnamefont {Togawa}},\ }\bibfield  {title} {\enquote {\bibinfo {title} {{Chirality-Induced Spin-Polarized State of a Chiral Crystal CrNb3 S6}},}\ }\href@noop {} {\bibfield  {journal} {\bibinfo  {journal} {Physical Review Letters}\ }\textbf {\bibinfo {volume}
  {124}},\ \bibinfo {pages} {166602} (\bibinfo {year} {2020})}\BibitemShut {NoStop}%
\bibitem [{\citenamefont {Bian}\ \emph {et~al.}(2023)\citenamefont {Bian}, \citenamefont {Nakano}, \citenamefont {Miyata}, \citenamefont {Oya}, \citenamefont {Nobuoka}, \citenamefont {Tsutsui}, \citenamefont {Seki},\ and\ \citenamefont {Suda}}]{Bian2023}%
  \BibitemOpen
  \bibfield  {author} {\bibinfo {author} {\bibfnamefont {Z.}~\bibnamefont {Bian}}, \bibinfo {author} {\bibfnamefont {Y.}~\bibnamefont {Nakano}}, \bibinfo {author} {\bibfnamefont {K.}~\bibnamefont {Miyata}}, \bibinfo {author} {\bibfnamefont {I.}~\bibnamefont {Oya}}, \bibinfo {author} {\bibfnamefont {M.}~\bibnamefont {Nobuoka}}, \bibinfo {author} {\bibfnamefont {Y.}~\bibnamefont {Tsutsui}}, \bibinfo {author} {\bibfnamefont {S.}~\bibnamefont {Seki}}, \ and\ \bibinfo {author} {\bibfnamefont {M.}~\bibnamefont {Suda}},\ }\bibfield  {title} {\enquote {\bibinfo {title} {{Chiral Van Der Waals Superlattices for Enhanced Spin‐Selective Transport and Spin‐Dependent Electrocatalytic Performance}},}\ }\href@noop {} {\bibfield  {journal} {\bibinfo  {journal} {Advanced Materials}\ }\textbf {\bibinfo {volume} {35}},\ \bibinfo {pages} {2306061} (\bibinfo {year} {2023})}\BibitemShut {NoStop}%
\bibitem [{\citenamefont {Menichetti}\ \emph {et~al.}(2023)\citenamefont {Menichetti}, \citenamefont {Cavicchi}, \citenamefont {Lucchesi}, \citenamefont {Taddei}, \citenamefont {Iannaccone}, \citenamefont {Jarillo-Herrero}, \citenamefont {Felser}, \citenamefont {Koppens},\ and\ \citenamefont {Polini}}]{Menichetti2024}%
  \BibitemOpen
  \bibfield  {author} {\bibinfo {author} {\bibfnamefont {G.}~\bibnamefont {Menichetti}}, \bibinfo {author} {\bibfnamefont {L.}~\bibnamefont {Cavicchi}}, \bibinfo {author} {\bibfnamefont {L.}~\bibnamefont {Lucchesi}}, \bibinfo {author} {\bibfnamefont {F.}~\bibnamefont {Taddei}}, \bibinfo {author} {\bibfnamefont {G.}~\bibnamefont {Iannaccone}}, \bibinfo {author} {\bibfnamefont {P.}~\bibnamefont {Jarillo-Herrero}}, \bibinfo {author} {\bibfnamefont {C.}~\bibnamefont {Felser}}, \bibinfo {author} {\bibfnamefont {F.~H.~L.}\ \bibnamefont {Koppens}}, \ and\ \bibinfo {author} {\bibfnamefont {M.}~\bibnamefont {Polini}},\ }\href@noop {} {\enquote {\bibinfo {title} {Giant chirality-induced spin polarization in twisted transition metal dichalcogenides},}\ } (\bibinfo {year} {2023}),\ \bibinfo {note} {preprint, doi: 10.48550/arXiv.2312.09169V2 (Accessed February 5, 2023)}\BibitemShut {NoStop}%
\bibitem [{\citenamefont {Michaeli}\ and\ \citenamefont {Naaman}(2019)}]{Michaeli2019a}%
  \BibitemOpen
  \bibfield  {author} {\bibinfo {author} {\bibfnamefont {K.}~\bibnamefont {Michaeli}}\ and\ \bibinfo {author} {\bibfnamefont {R.}~\bibnamefont {Naaman}},\ }\bibfield  {title} {\enquote {\bibinfo {title} {{Origin of Spin-Dependent Tunneling Through Chiral Molecules}},}\ }\href@noop {} {\bibfield  {journal} {\bibinfo  {journal} {Journal of Physical Chemistry C}\ }\textbf {\bibinfo {volume} {123}},\ \bibinfo {pages} {17043--17048} (\bibinfo {year} {2019})}\BibitemShut {NoStop}%
\bibitem [{\citenamefont {Das}, \citenamefont {Naaman},\ and\ \citenamefont {Fransson}(2024)}]{Das2024}%
  \BibitemOpen
  \bibfield  {author} {\bibinfo {author} {\bibfnamefont {T.~K.}\ \bibnamefont {Das}}, \bibinfo {author} {\bibfnamefont {R.}~\bibnamefont {Naaman}}, \ and\ \bibinfo {author} {\bibfnamefont {J.}~\bibnamefont {Fransson}},\ }\bibfield  {title} {\enquote {\bibinfo {title} {{Insights into the Mechanism of Chiral‐Induced Spin Selectivity: The Effect of Magnetic Field Direction and Temperature}},}\ }\href@noop {} {\bibfield  {journal} {\bibinfo  {journal} {Advanced Materials}\ }\textbf {\bibinfo {volume} {36}},\ \bibinfo {pages} {2313708} (\bibinfo {year} {2024})}\BibitemShut {NoStop}%
\bibitem [{\citenamefont {Moharana}\ \emph {et~al.}(2024)\citenamefont {Moharana}, \citenamefont {Kapon}, \citenamefont {Kammerbauer}, \citenamefont {Anthofer}, \citenamefont {Yochelis}, \citenamefont {Kläui}, \citenamefont {Paltiel},\ and\ \citenamefont {Wittmann}}]{Moharana2024}%
  \BibitemOpen
  \bibfield  {author} {\bibinfo {author} {\bibfnamefont {A.}~\bibnamefont {Moharana}}, \bibinfo {author} {\bibfnamefont {Y.}~\bibnamefont {Kapon}}, \bibinfo {author} {\bibfnamefont {F.}~\bibnamefont {Kammerbauer}}, \bibinfo {author} {\bibfnamefont {D.}~\bibnamefont {Anthofer}}, \bibinfo {author} {\bibfnamefont {S.}~\bibnamefont {Yochelis}}, \bibinfo {author} {\bibfnamefont {M.}~\bibnamefont {Kläui}}, \bibinfo {author} {\bibfnamefont {Y.}~\bibnamefont {Paltiel}}, \ and\ \bibinfo {author} {\bibfnamefont {A.}~\bibnamefont {Wittmann}},\ }\href@noop {} {\enquote {\bibinfo {title} {Chiral-induced unidirectional spin-to-charge conversion},}\ } (\bibinfo {year} {2024}),\ \bibinfo {note} {preprint, doi: 10.48550/arXiv.2402.19246V1 (Accessed April 1, 2023)}\BibitemShut {NoStop}%
\bibitem [{\citenamefont {Al-Bustami}\ \emph {et~al.}(2022)\citenamefont {Al-Bustami}, \citenamefont {Khaldi}, \citenamefont {Shoseyov}, \citenamefont {Yochelis}, \citenamefont {Killi}, \citenamefont {Berg}, \citenamefont {Gross}, \citenamefont {Paltiel},\ and\ \citenamefont {Yerushalmi}}]{AlBustami2022a}%
  \BibitemOpen
  \bibfield  {author} {\bibinfo {author} {\bibfnamefont {H.}~\bibnamefont {Al-Bustami}}, \bibinfo {author} {\bibfnamefont {S.}~\bibnamefont {Khaldi}}, \bibinfo {author} {\bibfnamefont {O.}~\bibnamefont {Shoseyov}}, \bibinfo {author} {\bibfnamefont {S.}~\bibnamefont {Yochelis}}, \bibinfo {author} {\bibfnamefont {K.}~\bibnamefont {Killi}}, \bibinfo {author} {\bibfnamefont {I.}~\bibnamefont {Berg}}, \bibinfo {author} {\bibfnamefont {E.}~\bibnamefont {Gross}}, \bibinfo {author} {\bibfnamefont {Y.}~\bibnamefont {Paltiel}}, \ and\ \bibinfo {author} {\bibfnamefont {R.}~\bibnamefont {Yerushalmi}},\ }\bibfield  {title} {\enquote {\bibinfo {title} {{Atomic and Molecular Layer Deposition of Chiral Thin Films Showing up to 99{\%} Spin Selective Transport}},}\ }\href@noop {} {\bibfield  {journal} {\bibinfo  {journal} {Nano letters}\ }\textbf {\bibinfo {volume} {22}},\ \bibinfo {pages} {5022--5028} (\bibinfo {year} {2022})}\BibitemShut {NoStop}%
\bibitem [{\citenamefont {Yang}, \citenamefont {{Van Der Wal}},\ and\ \citenamefont {{Van Wees}}(2019)}]{Yang2019}%
  \BibitemOpen
  \bibfield  {author} {\bibinfo {author} {\bibfnamefont {X.}~\bibnamefont {Yang}}, \bibinfo {author} {\bibfnamefont {C.}~\bibnamefont {{Van Der Wal}}}, \ and\ \bibinfo {author} {\bibfnamefont {B.}~\bibnamefont {{Van Wees}}},\ }\bibfield  {title} {\enquote {\bibinfo {title} {{Spin-dependent electron transmission model for chiral molecules in mesoscopic devices}},}\ }\href@noop {} {\bibfield  {journal} {\bibinfo  {journal} {Physical Review B}\ }\textbf {\bibinfo {volume} {99}},\ \bibinfo {pages} {024418} (\bibinfo {year} {2019})}\BibitemShut {NoStop}%
\bibitem [{\citenamefont {Tirion}\ and\ \citenamefont {van Wees}(2024)}]{Tirion2024}%
  \BibitemOpen
  \bibfield  {author} {\bibinfo {author} {\bibfnamefont {S.~H.}\ \bibnamefont {Tirion}}\ and\ \bibinfo {author} {\bibfnamefont {B.~J.}\ \bibnamefont {van Wees}},\ }\bibfield  {title} {\enquote {\bibinfo {title} {{Mechanism for Electrostatically Generated Magnetoresistance in Chiral Systems without Spin-Dependent Transport}},}\ }\href@noop {} {\bibfield  {journal} {\bibinfo  {journal} {ACS Nano}\ }\textbf {\bibinfo {volume} {18}},\ \bibinfo {pages} {6028--6037} (\bibinfo {year} {2024})}\BibitemShut {NoStop}%
\bibitem [{\citenamefont {Xiao}, \citenamefont {Zhao},\ and\ \citenamefont {Yan}(2023)}]{xiao2023nonreciprocal}%
  \BibitemOpen
  \bibfield  {author} {\bibinfo {author} {\bibfnamefont {J.}~\bibnamefont {Xiao}}, \bibinfo {author} {\bibfnamefont {Y.}~\bibnamefont {Zhao}}, \ and\ \bibinfo {author} {\bibfnamefont {B.}~\bibnamefont {Yan}},\ }\href@noop {} {\enquote {\bibinfo {title} {Nonreciprocal nature and induced tunneling barrier modulation in chiral molecular devices},}\ } (\bibinfo {year} {2023}),\ \bibinfo {note} {preprint, doi: 10.48550/arXiv.2201.03623V3 (Accessed August 5, 2023)}\BibitemShut {NoStop}%
\bibitem [{\citenamefont {Das}\ \emph {et~al.}(2018)\citenamefont {Das}, \citenamefont {Jousma}, \citenamefont {Majumdar},\ and\ \citenamefont {Banerjee}}]{Das2018}%
  \BibitemOpen
  \bibfield  {author} {\bibinfo {author} {\bibfnamefont {A.}~\bibnamefont {Das}}, \bibinfo {author} {\bibfnamefont {S.~T.}\ \bibnamefont {Jousma}}, \bibinfo {author} {\bibfnamefont {A.}~\bibnamefont {Majumdar}}, \ and\ \bibinfo {author} {\bibfnamefont {T.}~\bibnamefont {Banerjee}},\ }\bibfield  {title} {\enquote {\bibinfo {title} {{Evolution of the magnetoresistance lineshape with temperature and electric field across Nb-doped SrTiO3 interface}},}\ }\href@noop {} {\bibfield  {journal} {\bibinfo  {journal} {Applied Physics Letters}\ }\textbf {\bibinfo {volume} {112}},\ \bibinfo {pages} {3--6} (\bibinfo {year} {2018})}\BibitemShut {NoStop}%
\bibitem [{\citenamefont {Korytár}, \citenamefont {van Ruitenbeek},\ and\ \citenamefont {Evers}(2024)}]{Korytár2024}%
  \BibitemOpen
  \bibfield  {author} {\bibinfo {author} {\bibfnamefont {R.}~\bibnamefont {Korytár}}, \bibinfo {author} {\bibfnamefont {J.~M.}\ \bibnamefont {van Ruitenbeek}}, \ and\ \bibinfo {author} {\bibfnamefont {F.}~\bibnamefont {Evers}},\ }\href@noop {} {\enquote {\bibinfo {title} {Spin conductances and magnetization production in chiral molecular junctions},}\ } (\bibinfo {year} {2024}),\ \bibinfo {note} {preprint, doi: 10.48550/arXiv.2404.05614V4 (Accessed August 29, 2023)}\BibitemShut {NoStop}%
\bibitem [{\citenamefont {Jacquod}\ \emph {et~al.}(2012)\citenamefont {Jacquod}, \citenamefont {Whitney}, \citenamefont {Meair},\ and\ \citenamefont {B{\"{u}}ttiker}}]{Jacquod2012}%
  \BibitemOpen
  \bibfield  {author} {\bibinfo {author} {\bibfnamefont {P.}~\bibnamefont {Jacquod}}, \bibinfo {author} {\bibfnamefont {R.~S.}\ \bibnamefont {Whitney}}, \bibinfo {author} {\bibfnamefont {J.}~\bibnamefont {Meair}}, \ and\ \bibinfo {author} {\bibfnamefont {M.}~\bibnamefont {B{\"{u}}ttiker}},\ }\bibfield  {title} {\enquote {\bibinfo {title} {{Onsager relations in coupled electric, thermoelectric, and spin transport: The tenfold way}},}\ }\href@noop {} {\bibfield  {journal} {\bibinfo  {journal} {Physical Review B}\ }\textbf {\bibinfo {volume} {86}},\ \bibinfo {pages} {246604} (\bibinfo {year} {2012})}\BibitemShut {NoStop}%
\bibitem [{\citenamefont {Wolf}\ \emph {et~al.}(2022)\citenamefont {Wolf}, \citenamefont {Liu}, \citenamefont {Xiao}, \citenamefont {Park},\ and\ \citenamefont {Yan}}]{Wolf2022}%
  \BibitemOpen
  \bibfield  {author} {\bibinfo {author} {\bibfnamefont {Y.}~\bibnamefont {Wolf}}, \bibinfo {author} {\bibfnamefont {Y.}~\bibnamefont {Liu}}, \bibinfo {author} {\bibfnamefont {J.}~\bibnamefont {Xiao}}, \bibinfo {author} {\bibfnamefont {N.}~\bibnamefont {Park}}, \ and\ \bibinfo {author} {\bibfnamefont {B.}~\bibnamefont {Yan}},\ }\bibfield  {title} {\enquote {\bibinfo {title} {{Unusual Spin Polarization in the Chirality-Induced Spin Selectivity}},}\ }\href@noop {} {\bibfield  {journal} {\bibinfo  {journal} {ACS Nano}\ }\textbf {\bibinfo {volume} {16}},\ \bibinfo {pages} {18601--18607} (\bibinfo {year} {2022})}\BibitemShut {NoStop}%
\bibitem [{\citenamefont {Jansen}\ \emph {et~al.}(2012)\citenamefont {Jansen}, \citenamefont {Dash}, \citenamefont {Sharma},\ and\ \citenamefont {Min}}]{Jansen2012}%
  \BibitemOpen
  \bibfield  {author} {\bibinfo {author} {\bibfnamefont {R.}~\bibnamefont {Jansen}}, \bibinfo {author} {\bibfnamefont {S.~P.}\ \bibnamefont {Dash}}, \bibinfo {author} {\bibfnamefont {S.}~\bibnamefont {Sharma}}, \ and\ \bibinfo {author} {\bibfnamefont {B.~C.}\ \bibnamefont {Min}},\ }\bibfield  {title} {\enquote {\bibinfo {title} {{Silicon spintronics with ferromagnetic tunnel devices}},}\ }\href@noop {} {\bibfield  {journal} {\bibinfo  {journal} {Semiconductor Science and Technology}\ }\textbf {\bibinfo {volume} {27}},\ \bibinfo {pages} {083001} (\bibinfo {year} {2012})}\BibitemShut {NoStop}%
\bibitem [{\citenamefont {Fert}\ and\ \citenamefont {Jaffr{\`{e}}s}(2001)}]{Fert2001}%
  \BibitemOpen
  \bibfield  {author} {\bibinfo {author} {\bibfnamefont {A.}~\bibnamefont {Fert}}\ and\ \bibinfo {author} {\bibfnamefont {H.}~\bibnamefont {Jaffr{\`{e}}s}},\ }\bibfield  {title} {\enquote {\bibinfo {title} {{Conditions for efficient spin injection from a ferromagnetic metal into a semiconductor}},}\ }\href@noop {} {\bibfield  {journal} {\bibinfo  {journal} {Physical Review B}\ }\textbf {\bibinfo {volume} {64}},\ \bibinfo {pages} {184420} (\bibinfo {year} {2001})}\BibitemShut {NoStop}%
\end{thebibliography}%

\end{document}